\documentclass[12pt]{article}
\usepackage{graphicx}
\date{}
\def\be{\begin{equation}}
\def\ee{\end{equation}}
\def\bea{\begin{eqnarray}}
\def\eea{\end{eqnarray}}
\def\s{\sigma}

\def\om{\omega}
\def\pr{\prime}
\def\f{\varphi}
\def\th{\theta}

\def\tom{\tilde\omega}

\title{UNSTABLE STATES FOR CLOSED STRING \\
WITH MASSIVE POINT\thanks{Is supported
by Russian foundation of basic research, grant 05-02-16722}\\
}

\author{G.\,S. Sharov\\
{\small Tver state university}\\
{\small Tver, 170002, Sadovyj per. 35, Mathem. dep-t}}
\begin{document}
\maketitle
\begin{abstract}
The stability problem for the hypocycloidal rotational states of the closed
relativistic string with a point-like mass is solved with the help of
analysis of small disturbances of these states. Both analytical and
numerical investigations showed an unexpected result: the mentioned states
turned out to be unstable. This conclusion is based upon the presence of roots
with positive imaginary parts (increments) in the spectrum of frequencies
of small disturbances.
But these increments were small enough, so this instability had not
been detected in previous numerical experiments.
For the linear rotational states (the particular case of hypocycloidal states)
the stability was confirmed.
These results are important for applications of this model
in hadron spectroscopy.
\end{abstract}

\centerline {\bf 1. Introduction}
\medskip

In the previous work \cite{clstab05} we investigated the stability problem
for 3 classes of rotational states of the closed string with a point-like mass.
The corresponding exact solutions of the dynamical equations for this string
were obtained in Ref.~\cite{MilSh}. The closed string carrying one massive
point moves in the space ${\cal M}=R^{1,3}\times T^{D-4}$, where
$R^{1,3}$ is Minkowski space, $T^{D-4}$ is torus resulting from
the compactification procedure \cite{GSW}.
We denote $x^0,\dots,x^3$ the coordinates
in $R^{1,3}$ and the torus $T^{D-4}$ has
cyclic coordinates $x^k$ ($k=4,5\dots$) with periods $\ell_k$,
that is, points with coordinates $x^k$ and $x^k+N_k\ell_k$,
$N_k\in Z$ are identified. The metric in ${\cal M}$ is flat one:
$ds^2=\eta_{\mu\nu}dx^\mu dx^\nu$, the corresponding basis
$e_0,\;e_1,\;e_2,\,\dots\, e_{D-1}$ is orthonormal one.

For two (from 3) classes of the mentioned rotational states in
Ref.~\cite{clstab05} we used two approaches of solving the stability problem:
1) numerical simulation of motions, close to the rotational ones,
and 2) analytical investigation of small disturbances spectra
for these states.
The both approaches gave the same results. But for the third class ---
hypocycloidal rotational states we used only numerical simulation
because of very sophisticated calculations in the spectral analysis.
But later, when the calculations had been fulfilled, the
unexpected results were obtained. They are presented in this paper.

\bigskip

\centerline {\bf 2. Dynamics}
\medskip

The dynamics of the closed string (with the world surfaces $X^\mu(\tau,\s)$,
$\s_1\le\s\le\s_2)$ carrying a point-like mass $m$ in the
space ${\cal M}$ is described by the system of equations  \cite{MilSh}
\be
\frac{\partial^2X^\mu}{\partial\tau^2}-
\frac{\partial^2X^\mu}{\partial\s^2}=0,
\label{eq}\ee
\vspace{-3mm}
\be
X^\mu(\tau^*,2\pi)=X^\mu(\tau,0)+\sum\limits_{k\ge4}N_k\ell_k e^\mu_k
\label{clos}\ee
\vspace{-3mm}
\be
m\frac d{d\tau}\frac{\dot X^\mu(\tau,0)}{\sqrt{\dot X^2(\tau,0)}}+\gamma
\big[X^{'\!\mu}(\tau^*,2\pi)-X^{'\!\mu}(\tau,0)\big]=0,
\label{mg}\ee
which (without loss of generality) take this form under the conditions
$\s_1=0$, $\s_2=2\pi$ and the orthonormality conditions on the world surface
\be
(\partial_\tau X\pm\partial_\s X)^2=0.
\label{ort}\ee
Here $\gamma$ is the string tension, $\dot X^\mu\equiv\partial_\tau X^\mu $,
$X^{'\!\mu}\equiv\partial_\s X^\mu$; the scalar product is
$(a,b)=\eta_{\mu\nu}a^\mu b^\nu $.

Equation (\ref{clos}) is the closure condition on the tube-like world surface
of the closed sting on the world line of the massive point \cite{PRTr}.
This line can be parameterized with two different parameters $\tau$ and
$\tau^*$, connected by the relation $\tau^*=\tau^*(\tau).$
This relation should be added to the closure condition (\ref{clos}).

In this paper we consider the hypocycloidal rotational states corresponding
to the following solutions of the system (\ref{eq})\,--\,(\ref{ort})
\cite{clstab05}, \cite{MilSh}:
\be
\begin{array}{c}\displaystyle
X^\mu(\tau,\s)=e_0^\mu a_0(\tau-\th\s)+\sum\limits_{k>3}e^\mu_k b_k\s+\\
+A\Big\{\big[S\cos\om\s+(C_\th-C)\sin\om\s\big]\cdot
e^\mu(\om\tau)-S_\th\sin\om\s\cdot\acute e^\mu(\om\tau)
\Big\}.
\rule[3mm]{0mm}{1mm}
\end{array}
\label{hyp}\ee
Here $\displaystyle b_k =\frac {\ell_k N_k}{2\pi}$,
 $e^\mu(\om\tau)=e^\mu_1\cos\om\tau+e^\mu_2\sin\om\tau$,

$\acute e^\mu(\om\tau)=\om^{-1}\frac d{d\tau} e^\mu(\om\tau)$
are unit orthogonal rotating vectors, the speed of light $c=1$,
the following values
\be
Q=\frac\gamma{m}\sqrt{\dot X^2(\tau,0)}={}\mbox{const},\qquad
\th=\frac{\tau^*(\tau)-\tau}{2\pi}={}\mbox{const},
\label{Qth}\ee
are constant for sulutions (\ref{hyp}),
\be
C =\cos2\pi\om, \;\quad S=\sin2\pi\om,\;
\quad C_\th =\cos2\pi\th\om, \;\quad S_\th=\sin2\pi\th\om.
\label{CS}
\ee
Values $a_0$, $A$ and speed of the massive point $v={}$const
are connected by the equations
\be
a_0 =\frac{mQ}{\gamma\sqrt{1-v^2}}, \qquad
A =\frac{a_0v}{\om S},\qquad v^2 =\th\frac{S}{S_\th},
\label{a0A}
\ee
and values $\om$ and $\th$ are determined from the system  \cite{MilSh}
\be
C-C_\th= \frac{\om}{2Q}S,
\label{con}\ee
\vspace{-3mm}
\be
(1+\th^2)\,SS_\th-2\th(1-CC_\th)=
(S_\th-\th S)\frac{\beta S}{Q^2},
\label{omtaub}\ee
resulting from Eqs.~(\ref{eq})\,--\,(\ref{ort}).
Here $\beta=(\gamma/m)^2\sum\limits_{k>3}b_k^2$

Solution of the system (\ref{con}), (\ref{omtaub}) (pairs $\om$, $\th$)
form some countable set. Each pair corresponds to solution (\ref{hyp})
describing uniform rotation of the closed string with certain topological
type. In the case $\beta=0$ the string has the form of a closed
hypocycloid joined at non-zero angle in the massive point, so we
use the term ``hypocycloidal rotational states'' for motions (\ref{hyp}).
These states generate
non-trivial spectrum of Regge trajectories \cite{MilSh} and may be
applied in the hadron spectroscopy.

Structure of solutions of the system (\ref{con}), (\ref{omtaub})
is illustrated in Fig.~1. Here solid lines are determined by Eq.~(\ref{con})
and dashed lines form the graphical chart of Eq.~(\ref{omtaub}) in the
($\om,\th$) plane. Here $Q=1$, $\beta=0.1$.  Solutions ($\om,\th$)
of this system are connected to cross points of these lines.

Note that solutions $\om=n/2$, $\th=(n-2k)/n$ of Eqs.~(\ref{con}),
(\ref{omtaub}) correspond to hypocycloidal states (\ref{hyp}) of
the closed string with zero mass $m=0$. But other solutions (cross
points in Fig.~1) result in states with $m\ne0$. Projection of the
string onto $e_1,\,e_2$ plane is a curvilinear $n$-gon (a
hypocycloid in the case $\beta=0$), where $n$ is the number of
solid line in Fig.~1.

\begin{figure}[th]
\unitlength=1pt \centering
\includegraphics[width=\textwidth]{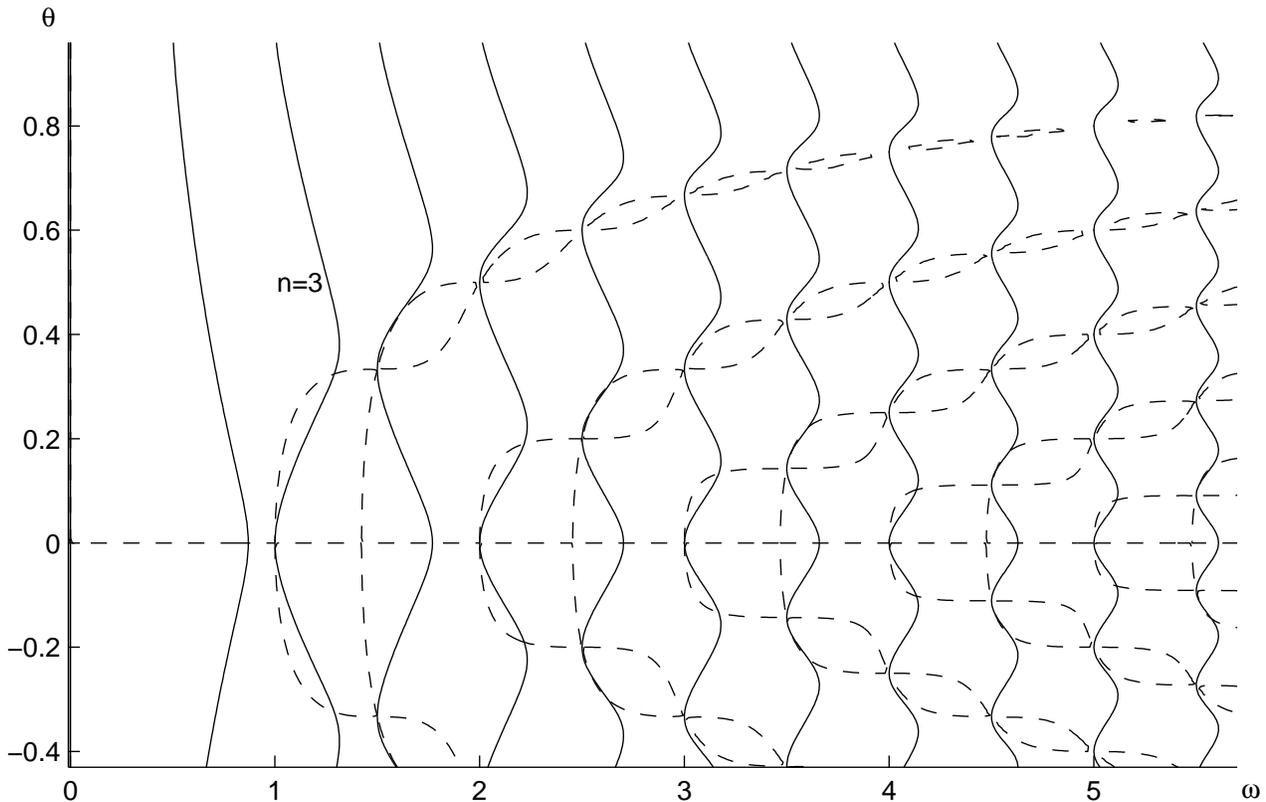}
\caption{\small Graphical chart of Eq.~(\ref{con}) (solid lines)
and  Eq.~(\ref{omtaub}) (dashed lines) for $Q=1$, $\beta=0.1$ }
\end{figure}

In the particular case $\th=0$ solutions (\ref{hyp}) with frequencies
$\om$ (roots of Eq.~(\ref{con}), precisely, of equation $\om=-2Q\tan\pi\om$)
have the form
\be X^\mu=e_0^\mu a_0\tau
+\sum_{k>3}e^\mu_k b_k\s+A\cos\big[\om(\s-\pi)\big]\cdot e^\mu (\om\tau).
\label{rot}
\ee
They describe uniform rotations of the sinusoidal string with rotating
(along a circle) massive point. We name these motions as
``linear rotational states'', because projections of the string onto
$e_1,\,e_2$ plane are rectilinear segments.

The third class of rotational states (with massive point at the center
of rotation) was studied in detail in Ref.~\cite{clstab05}.

Possible applications of solutions (\ref {hyp}) and (\ref {rot})
in hadron spectroscopy essentially depend on stability
or instability of these states with respect to small disturbances.
In the following section we study spectrum of these disturbances.

\bigskip

\centerline {\bf 3. Spectrum of disturbances}
\medskip

To solve the stability problem for rotational motions
(\ref{hyp}), (\ref{rot})
we consider the general solution of Eq.~(\ref{eq})
\be
X^\mu (\tau,\s)=\frac{1}{2}[\Psi^\mu_{+}(\tau +\s)+\Psi
^\mu_{-}(\tau-\s)]
\label{gensol}\ee
and denote
\be
\breve\Psi^{\pr\mu}_\pm (\tau)=e_0^\mu(1\mp\th)\,
a_0\pm\sum_{k>3} e^\mu_k b_k+\om A\Big[\pm(C_\th-C)\,e ^\mu(\om\tau)+
(S_\th\mp S)\,\acute e^\mu(\om\tau)\Big]
\label{Psihyp} \ee
the functions in the expression (\ref {gensol}) for
the considered rotational motions (\ref {hyp}).

To describe any small disturbances of the rotational motion of the
system, that is motions close to states (\ref{hyp})
we consider vector functions $\Psi^{\pr\mu}_\pm$
close to $\breve\Psi^{\pr\mu}_\pm$ in the form
\be
\Psi^{\pr\mu}_\pm(\tau)=\breve\Psi^{\pr\mu}_\pm(\tau)
+\f_\pm^\mu(\tau).\label{Psi+f}\ee

The disturbance $\f_\pm^\mu(\tau)$ is supposed to be small, so we
omit squares of $\f_\pm$ when we substitute the expression (\ref{Psi+f})
into dynamical equations (\ref{clos}) and (\ref {mg}).
In other words, we work in the first linear vicinity of the states
(\ref{hyp}).
Both functions $\Psi^{\pr\mu}_\pm$ and $\breve\Psi^{\pr\mu}_\pm$ in expression
(\ref{Psi+f}) must satisfy the condition
${\Psi'_+\!\!}^2={\Psi'_-\!\!}^2=0$, resulting from Eq.~ (\ref{ort}),
hence in the first order
approximation on $\f_\pm$ the following scalar product equals zero:
\be
\big\langle\breve\Psi^\pr_\pm,\f_\pm\big\rangle=0.
\label {Psif}
\ee

For the disturbed motions the equality (\ref{Qth}) $\tau^*=\tau+2\pi\th$,
generally speaking, is not carried out and should be replaced with the
equality
\be
\tau^*=\tau+2\pi\th+\zeta(\tau),
\label{taudel}\ee
where $\zeta(\tau)$ is a small disturbance.

Expression (\ref{Psi+f}) together with Eq.~(\ref{gensol}) is the
solution of the equations of string motion (\ref {eq}).
Therefore we can obtain the equations of evolution of small disturbances
$\f_\pm^\mu(\tau)$, substituting expressions (\ref{Psi+f}) and (\ref{taudel})
with Eq.~(\ref{Psihyp}) in two other equations of motion
(\ref{clos}) and (\ref{mg}).
We take into account the nonlinear factor $\big[\dot X^2(\tau,0)\big]^{-1/2}$
and contributions from the disturbed argument $\tau^*$ (\ref{taudel}):
$$
\breve\Psi^{\pr\mu}_\pm(\tau^*\pm2\pi)\simeq\breve\Psi^{\pr\mu}_\pm(\pm)+
\breve\Psi^{\pr\pr\mu}_\pm(\pm)\,\zeta(\tau).
$$
Here and below $(\pm)\equiv(\tau+\tau_0\pm2\pi)$.

This substitution results in the following linearized system of equations
in linear (with respect to $\f_\pm^\mu$ and $\zeta$) approximation:
\be
\begin {array}{c}
\f_+^\mu(+)+\f_-^\mu(-)+2a_0\big[e_0^\mu+v\acute e^\mu(\om \tau)\big]\,
\zeta'(\tau)-2a_0v\om  e^\mu(\om \tau)\,\zeta(\tau)=
\f_+^\mu(\tau)+\f_-^\mu(\tau),\\
\displaystyle
\frac d{d\tau}\bigg\{\f_+^\mu(\tau)+\f_-^\mu(\tau)-\frac1{1-v^2}
\big[e_0^\mu+v\acute e^\mu(\om \tau)\big]
\big[\f_+^0+\f_-^0+v(\acute\f_++\acute\f_-)\big]\bigg\}+
\rule{0mm}{7.5mm}\\
+Q\Big[\f_+^\mu(+)-\f_-^\mu(-)-\f_+^\mu(\tau)+\f_-^\mu(\tau)+
2\om ^2A\big[(C-C_\th)\,\acute e^\mu+S_\th e^\mu\big]\zeta(\tau)\Big]=0.
\rule{0mm}{6.5mm}\end{array}
\label{sysf}\ee
Here we use the following notations for the scalar products:
\be
\f_\pm^0\equiv\langle e_0,\f_\pm\rangle,\quad
\f_\pm^k\equiv\langle e_k,\f_\pm\rangle,\quad
\f_\pm\equiv \langle e,\f_\pm\rangle, \quad
\acute\f_\pm\equiv\langle\acute e,\f_\pm\rangle.
\label{fiscal}\ee

Projections (scalar products) of equations (\ref{sysf}) onto
vectors $e_0$, $e_k$, $e(\tau)$, $\acute e(\tau)$ form
the following system of equations:
\be
\begin {array}{c}
\f_+^k(+)+\f_-^k(-)=\f_+^k(\tau)+\f_-^k(\tau),\\
\dot\f_+^k(\tau)+\dot\f_-^k(\tau)+
Q\big[\f_+^k(+)-\f_-^k(-)-\f_+^k(\tau)+\f_-^k(\tau)\big]=0,
\rule{0mm}{5mm}\end{array}
\label{sysfk}\ee
\vspace{-1mm}
\be
\!\!\begin {array}{c}
\f_+^0(+)+\f_-^0(-)-\f_+^0(\tau)-\f_-^0(\tau)+2a_0\dot\zeta(\tau)=0,\\
C_+\f_+(+)+C_-\f_-(-)-S_+\acute\f_+(+)-S_-\acute\f_-(-)-\f_+(\tau)-\f_-(\tau)
+2a_0v\om\zeta(\tau)=0,
\rule{0mm}{5mm}\\
C_+\acute\f_+(+)+C_-\acute\f_-(-)+S_+\f_+(+)+S_-\f_-(-)-\acute\f_+(\tau)-
\acute\f_-(\tau)-2a_0v\dot\zeta(\tau)=0,\rule{0mm}{5mm}\\
\frac v{1-v^2}\frac d{d\tau}\Big[v\f_+^0(\tau)+v\f_-^0(\tau)
+\acute\f_+(\tau)+\acute\f_-(\tau)\Big]=
Q\Big[f_+^0(+)-\f_-^0(-)-\f_+^0(\tau)+\f_-^0(\tau)\Big],\rule{0mm}{5mm}\\
\dot\f_++\dot\f_--\frac\om{1-v^2}\Big[v(\f_+^0+\f_-^0)+\acute\f_++\acute\f_-
\Big]+\rule{0mm}{5mm}\\
\!\!+Q\Big[C_+\f_+(+)-S_+\acute\f_+(+)-C_-\f_-(-)+S_-\acute\f_-(-)
-\f_+(\tau)+\f_-(\tau)-2a_0\om\frac{\th}v\zeta(\tau)\Big]=0,\rule{0mm}{4mm}\!\!\\
\om(\f_++\f_-)+\frac1{1-v^2}\frac d{d\tau}
\Big[v(\f_+^0+\f_-^0)+\acute\f_++\acute\f_-\Big]+\rule{0mm}{5mm}\\
\!\!+Q\Big[C_+\acute\f_+(+)+S_+\f_+(+)-C_-\acute\f_-(-)-S_-\f_-(-)
-\acute\f_+(\tau)+\acute\f_-(\tau)-a_0\om^2\frac vQ\zeta(\tau)\Big]=0.
\rule{0mm}{4mm}\!\!\end{array}\!\!\!\!
\label{sysf1}\ee
Here $C_\pm=\cos\big[2\pi\om(\th\pm1)\big]=C_\th C\mp S_\th S$,
$S_\pm=S_\th C\pm C_\th S$, two equations (\ref{sysfk}) are
projections of Eqs.~(\ref{sysf}) onto vectors $e_k$, $k=3,\,4,\dots$

We should add to this system equations (\ref{Psif})
 after substituting expressions (\ref{Psihyp})
\be
(1\mp\th)\,\f_\pm^0\pm\frac1{a_0}\sum_{k>3}b_k\f_\pm^k+
\frac vS\big[\pm(C_\th-C)\,\f_\pm+(S\mp S_\th)\,\acute\f_\pm\big]=0
\label{fi7}\ee

System (\ref{sysfk})\,--\,(\ref{fi7}) is the linear system of differential
equations with respect to projections (\ref{fiscal})
$\f_\pm^0(\tau)$, $\f_\pm^k(\tau)$, $\f_\pm(\tau)$, $\acute\f_\pm(\tau)$,
and the function $\zeta(\tau)$.
This system has constant coefficients but it also has deviating arguments
$(\pm)$ together with $(\tau)$.

We search solutions of this system in the form of harmonics
\be
\f_\pm^0=B_\pm^0 e^{-i\tom\tau},\quad \f_\pm^k=B_\pm^k e^{-i\tom\tau},
\f_\pm=B_\pm e^{-i\tom\tau},\quad \acute \f_\pm=\acute B_\pm e^{-i\tom\tau},
\quad 2a_0\zeta=\Delta e^{-i\tom\tau}.
\label{fexp} \ee

This substitution results in the linear homogeneous system of algebraic
equations with respect to the amplitudes of harmonics (\ref{fexp}).
Two equations of this system connected with Eqs.~(\ref{sysfk}) are
\be
B_+^kE_+^1+B_-^kE_-^1= 0, \qquad B_+^k (QE_+^1-i\tom)
=B_-^k (QE_-^1+i\tom),
\label{sysBk}\ee
where $E_\pm^1=\exp\big[-i2\pi(\th\pm1)\,\tom\big]-1$.
System (\ref{sysBk}) has non-trivial solutions if and only if $\tom$ is a
root of the equation
\be
\cos2\pi\tom-\cos2\pi\th\tom=\frac{\tom}{2Q}\sin2\pi\tom,
\label{contom}\ee

It coincides with Eq.~(\ref{con}), if $\om$ is substituted by $\tom$.
The spectrum of transversal (with respect to the $e_1,\,e_2$ plane)
small fluctuations of the string for states (\ref {hyp}) contains
frequencies $\tom$ which are roots of Eq.~(\ref{contom}).
All these frequencies are real numbers, therefore amplitudes of such
fluctuations do not grow with growth of time $t$.

Another picture takes place for disturbances concerning to the
$e_1,\,e_2$ plane.
Assuming that frequencies $\tom$ of these fluctuations are not roots
of Eq.~(\ref{contom}), we find for these modes $B_\pm^k=0$ and
for other amplitudes (\ref{fexp}) equations (\ref{sysf1}) and
(\ref{fi7}) result in the following system (after transforming):
\be
\begin{array}{c}
\tom(\acute B_++\acute B_-)+(\tom v-igQE^1_+)\,B_+^0+
(\tom v+igQE^1_-)\,B_-^0=0,\\
\displaystyle
h_+E^1_+\acute B_+-h_-E^1_-\acute B_+=
\Big(i\frac{\om v}\tom E^1_+-h^*_+E_+^c\Big)B_+^0+
\Big(i\frac{\om v}\tom E^1_-+h^*_-E_-^c\Big)B_-^0,\rule{0mm}{6.5mm}\\
(E_++1)\,\acute B_++(E_-+1)\,\acute B_-=(vE^1_++h^*_+S_+E_+)\,B_+^0+
(vE^1_+-h^*_-S_-E_-)\,B_-^0,\rule{0mm}{5mm}\\
\big[\om q-h_+(E^1_+-i\tom/Q)\big]\acute B_++
\big[\om q-h_-(E^1_-+i\tom/Q)\big]\acute B_-= \rule{0mm}{5mm} \\
\displaystyle
=\Big[i\frac{\om\th}{\tom v}E^1_+-v\om q+h^*_+\Big(E_+^c-i\frac\tom Q\Big)\Big]B_+^0+
\Big[i\frac{\om\th}{\tom v}E^1_--v\om q+h^*_-\Big(E_-^c+i\frac\tom Q\Big)\Big]B_-^0;
\rule{0mm}{6.5mm}\\
(i\tom q+E^1_++2\th/v^2)\,\acute B_++
(i\tom q-E^1_-+2\th/v^2)\,\acute B_-= \rule{0mm}{5mm} \\
=\big[E^*_++h^*_+(S_+E_++\om/Q)\big]B_+^0+
\big[E^*_-+h^*_-(S_-E_--\om/Q)\big]B_-^0,\rule{0mm}{5mm}\\
\tom\Delta=-i(E^1_+B_+^0+E^1_-B_-^0),\qquad
B_\pm=\pm(h^*_\pm B_\pm^0+h_\pm\acute B_\pm).\rule{0mm}{5mm}
\end{array}
\label{sysBh}\ee
Here we use notations
$$
E_\pm=\exp\big[-i2\pi(\th\pm1)\,\tom\big],\qquad E_\pm^1=E_\pm-1,\qquad
E_\pm^c=C_\pm E_\pm-1,\qquad E^*_\pm=i\frac{\om^2v}{2Q\tom}E^1_\pm-iv\tom q
$$
\vspace{-3mm}
$$
g=\frac{1-v^2}v,\qquad h_\pm=\frac{2Q}\om\Big(1\mp\frac\th{v^2}\Big)=
\frac{S\mp S_\th}{C-C_\th},\qquad h^*_\pm=\frac{2Q}\om\frac{1\mp\th}v,
\qquad q=\frac1{Q(1-v^2)},
$$
equations (\ref{CS})\,--\,(\ref{omtaub}) and relations
$$
C_\pm\pm h_\pm S_\pm=-1,\qquad h_\pm(C_\pm-1)=\pm S_\pm.
$$

Notice that the fifth equations of system (\ref{sysBh}) is linear
combination of the first three ones with coefficients
$iq$, $\om/(2Q)$ and $\th/v^2$.
Hence, the condition of existence of non-trivial solutions for this
system is vanishing the determinant connected with the first four
equations. This condition results in the following equation:
\be
\begin{array}{c}
\displaystyle
\bigg(D\frac{\tilde C-\tilde C_\th}{\tom^2}-4gQ\frac{\tilde S}{\tom}
+V\tilde C_\th-g\tilde C\bigg)
\bigg(\frac{\om^2v}{2Q}\frac{\tilde S}{\tom}+V\tilde C_\th+g\tilde C\bigg)+
f(\tom)+\\
\displaystyle
+i\frac{\om^2 v}{2Q\tom}\bigg[V\tilde S\tilde S_\th+V_\th
(\tilde C\tilde C_\th-1)+2gQ(\tilde C-\tilde C_\th)
\frac{S\tilde S_\th-S_\th\tilde S}{S\tom}\bigg]=0.
\rule{0mm}{6mm}
\end{array}
\label{oschyp}\ee

Here the notations like (\ref{CS}) are used, but with $\tom$ instead of
$\om$: $\tilde C =\cos2\pi\tom$, $\tilde S
=\sin2\pi\tom$, $\tilde C_\th=\cos2\pi\th\tom$, $\tilde S_\th
=\sin2\pi\th\tom$; and also
$$
D=\om^2(gh_+h_-+v^{-1}),\quad
V=\frac{1-CC_\th}{v}\cdot\frac{C-v^2C_\th}{C-C_\th},\quad
V_\th=\frac{SS_\th}{v}\cdot\frac{C-v^2C_\th}{C-C_\th};
$$
\vspace{-3mm}
$$
f(\tom)=\Big(V^2-V_\th^2+\frac{g\om^2v-4g^2Q^2-DV}{\tom^2}\Big)
\tilde S_\th^2+\Big[\frac{g\th}{v^2}(\om^2v+8gQ^2)+DV_\th\Big]
\frac{\tilde S\tilde S_\th}{\tom^2}+
$$
\vspace{-3mm}
$$
+S_\th\bigg[4gQ\Big(\frac Sv-v\frac{1-CC_\th}{S}\Big)-
\om(C-v^2C_\th)\bigg]\frac{\tilde C\tilde S_\th}{\tom}
-\Big(V_\th\tilde C_\th-\frac{2gQ\th}{v^2\tom}\tilde S\Big)^2.
$$

Equation (\ref{oschyp}) has the imaginary part.
One can expect that complex (or imaginary) roots $\tom=\tom_r+i\xi$
 of this equation exist.

But in the case $\th=0$ that is for linear rotational states (\ref{rot})
the mentioned imaginary part vanishes. In this case equalities
$S_\th=\tilde S_\th=V_\th=0$ take place, hence, the function $f(\tom)=0$
and equation (\ref{oschyp}) decomposes into product of the following
two equations:
\be
2Q\tom\Big[v^2-C+(1-v^2)\cos2\pi\tom\Big]+\om ^2v^2\sin2\pi\tom=0,
\label{osc1}\ee
\vspace{-3mm}
\be
\left(\frac{\om^2}v+4gQ^2\right)(1-\cos2\pi\tom)+
4gQ\tom\sin2\pi\tom+\tom^2\left(\frac{C-v^2}v+g\cos2\pi\tom\right)=0.
\label{osc2}\ee

Their roots were analyzed in Ref.~\cite{clstab05}.
It was shown that all roots of Eq.~(\ref{osc1}) and

Eq.~(\ref{osc2}) are real numbers, if all values satisfy natural physical
restrictions, for example, $v<1$, $m>0$. The typical picture of these roots
in the plane $\tom_r,\,\xi$ is presented in Fig.~1a for Eq.~(\ref{osc1})
and in Fig.~1b for
Eq.~(\ref{osc2}). Roots are points of intersection of zero level lines
for real part Re$F(\tom_r+i\xi)=0$ (solid lines)
or imaginary part (dashed lines) of the corresponding equation with
$\tom=\tom_r+i\xi$.

\begin{figure}[h]
\unitlength=1pt \centering
\includegraphics[width=\textwidth]{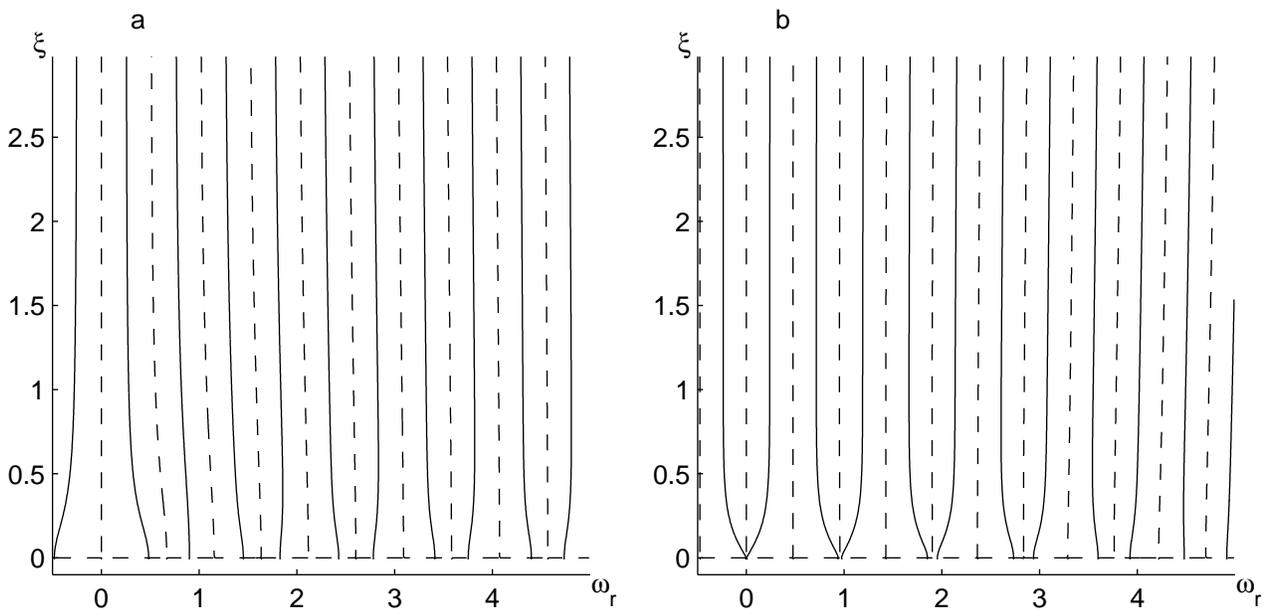}
\caption{\small Zero level lines for real (solid) and imaginary
part (dashed) for a) Eq.~(\ref{osc1}); b) Eq.~(\ref{osc2})}
\end{figure}

Here the values of the parameters for the linear rotational state (\ref{rot})
are: $\om =0.9$, $Q\simeq1.385$, $v^2\simeq0.875$,
$\beta=0.5$. We may conclude that in first order
approximation linear rotational states (\ref{rot}) are stable with
respect to small disturbances.

Let us turn to the hypocycloidal rotational states (\ref{hyp}) with
spectral equation (\ref{oschyp}) for small disturbances.

Substituting $\tom=\tom_r+i\xi$ into Eq.~(\ref{oschyp}) we draw in
Figs.~3,\,4 zero level lines for real part of l.\,h.\,s of this
equation (solid lines) and for its imaginary part (dashed lines)
similar to Fig.~2. In Fig.~3 these lines are shown for the
hypocycloidal state (\ref{hyp}) with the following values of
parameters: $Q=1$, $\beta=0{.}1$, $\om=1{.}289$, $\th=0{.}3242$.
The shape of the string (its projection onto $e_1,\,e_2$ plane) at
the corner of Fig.~3 is close to a curvilinear triangle.

\begin{figure}[h]
\unitlength=1pt \centering
\includegraphics[width=\textwidth]{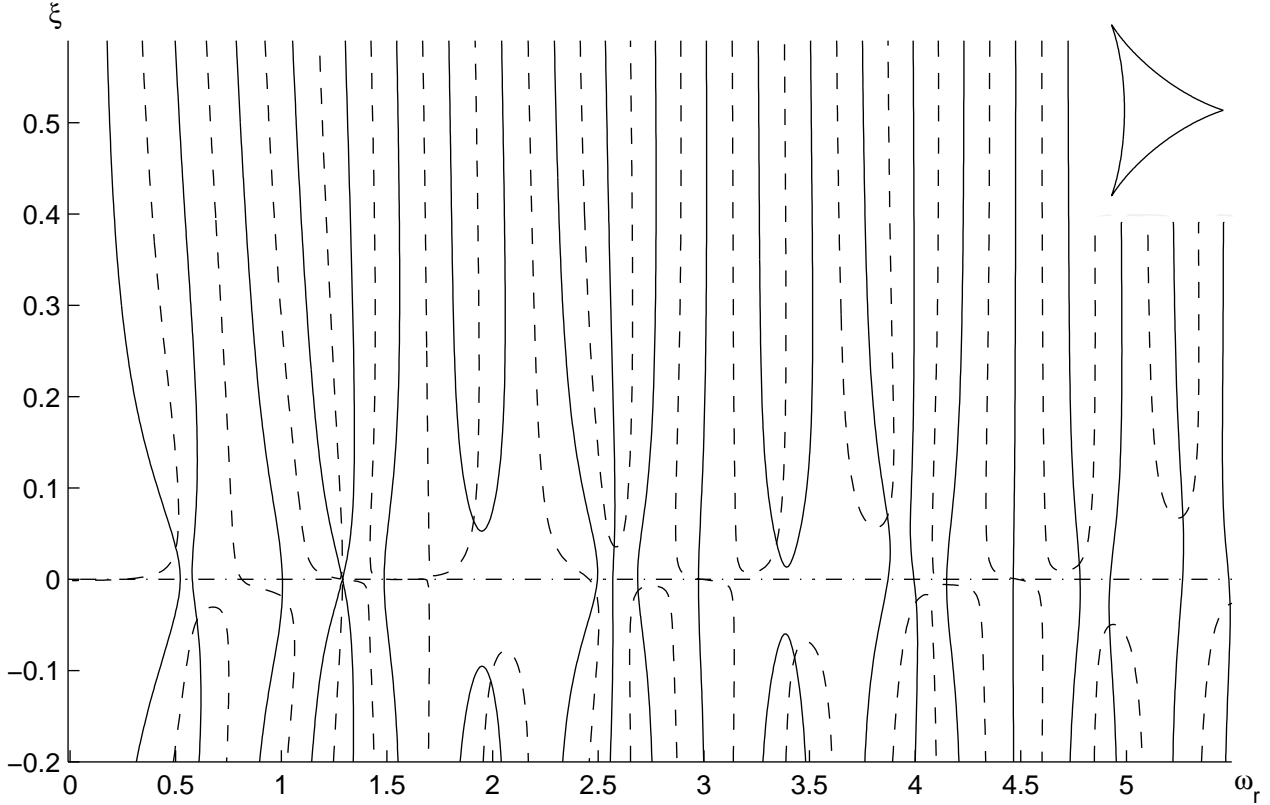}
\caption{\small Zero level lines for real (solid) and imaginary
part of Eq.~(\ref{oschyp}) for the ``triangle" state}
\end{figure}

In Fig.~4 the similar picture is drawn for the state
(\ref{hyp}) (hypocycloid with 4 arcs) for the recorded values of parameters.

\begin{figure}[h]
\unitlength=1pt \centering
\includegraphics[width=\textwidth]{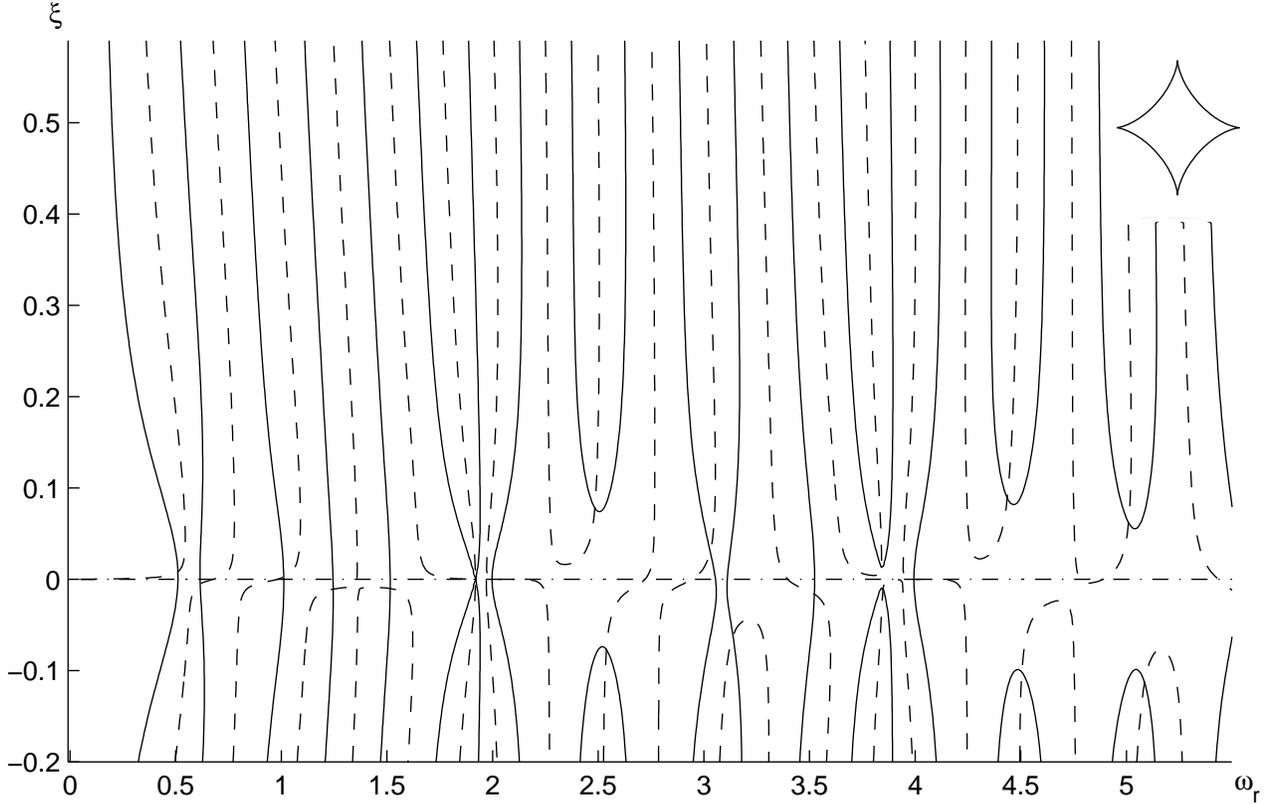}
\caption{\small Zero level lines for real (solid) and imaginary
part of Eq.~(\ref{oschyp}), $Q=5$, $\beta=0$, $\om=1{.}919$,
$\th=0{.}4996$ }
\end{figure}

In both cases in Figs.~3,\,4 one can find a set of cross points
(roots of Eq.~(\ref{oschyp})) $\tom=\tom_r+i\xi$ with positive
imaginary part $\xi>0$. The corresponding modes of
disturbances $\f^\mu$  get the
multiplier $\exp(\xi\tau)$, that is they grow exponentially.
This picture takes place for any (physically admissible) values of parameters
$Q$, $\beta$, $\om$, $\th$ so we may conclude that
the hypocycloidal rotational states (\ref{hyp}) for $\th>0$
are unstable with respect to small disturbances.

But maximal increments $\xi$ for growing amplitudes are small:
they do not exceed $0{.}1$ not only for the cases in Figs.~3,\,4, but also
for other topologically different hypocycloidal states with
various values $Q$, $\beta$, $\om$, $\th$.
So the growth factor for an amplitude of disturbance
$\exp(2\pi\xi/\om)$ (per one rotation) was not large enough
to detect this instability in previous numerical experiments
in Ref.~\cite{clstab05}. On should add that the maximal increments $\xi$
tend to zero in both limits $Q\to0$ (it corresponds to
$m\to\infty$ if $a_0\gamma={}$const) and $Q\to\infty$ ($m\to0$).

More detailed numerical simulation of string motions,
close to hypocycloidal states (\ref{hyp}) (slightly disturbed rotations)
shows that amplitudes of small disturbances grow with increments
$\xi$ described by Eq.~(\ref{oschyp}).

\bigskip

\centerline {\bf Conclusion}
\medskip

The analysis of stability for the hypocycloidal rotational states
(\ref{hyp}) of the closed string with a point-like mass demonstrated
instability of these states with $\th\ne0$ with respect to small disturbances.
It is similar to behavior of rotational states of Y string baryon model
\cite{stabPRD}, \cite{Y02}.
However, for the states (\ref{hyp}) increments $\xi$ of this
instability are small.
For the linear rotational states (\ref{rot}) (the particular case of
hypocycloidal states) the stability was confirmed.

This results are essential for possible applications of these states
for describing excited hadrons with exotic properties
(in particular, glueballs, hybrids, pentaquarks)
in accordance with applications of other string hadron models
\cite{4B} in meson and baryon spectroscopy.

\end{document}